\documentclass[aps,prb,reprint,amsmath,amssymb]{revtex4-2}

\usepackage{graphicx}
\usepackage{bm}
\usepackage[colorlinks,linkcolor=blue,citecolor=blue,urlcolor=blue]{hyperref}

\newcommand{\e}{\mathrm{e}}
\renewcommand{\i}{\mathrm{i}}
\renewcommand{\d}{\mathrm{d}}

\begin{document}

\title{Correlation effects in a simplified bilayer two-orbital Hubbard model at half filling}

\author{Jian-Jian Yang}
\affiliation{Guangdong Provincial Key Laboratory of Magnetoelectric Physics and Devices,\\
State Key Laboratory of Optoelectronic Materials and Technologies, Center for Neutron Science and Technology,\\
and School of Physics, Sun Yat-sen University, Guangzhou 510275, China}

\author{Dao-Xin Yao}
\email{yaodaox@mail.sysu.edu.cn}
\affiliation{Guangdong Provincial Key Laboratory of Magnetoelectric Physics and Devices,\\
State Key Laboratory of Optoelectronic Materials and Technologies, Center for Neutron Science and Technology,\\
and School of Physics, Sun Yat-sen University, Guangzhou 510275, China}

\author{Han-Qing Wu}
\email{wuhanq3@mail.sysu.edu.cn}
\affiliation{Guangdong Provincial Key Laboratory of Magnetoelectric Physics and Devices,\\
State Key Laboratory of Optoelectronic Materials and Technologies, Center for Neutron Science and Technology,\\
and School of Physics, Sun Yat-sen University, Guangzhou 510275, China}

\begin{abstract}
Motivated by the discovery of high-temperature superconductivity in bilayer nickelate La$_3$Ni$_2$O$_7$ under pressure, we investigate the ground-state phase diagram and correlation effects using determinant quantum Monte Carlo simulations in a simplified bilayer two-orbital Hubbard model at half filling.
Our results reveal the emergence of a nonmagnetic weakly insulating phase at weak on-site Hubbard interactions, transitioning to an antiferromagnetic Mott insulating phase as the interaction strength exceeds a critical value $U/t_1^x\approx4.15$.
This phase transition is consistent with the 3D O(3) Heisenberg universality class.
Additionally, we analyze dynamical properties such as the single-particle spectral function and dynamic spin structure factor.
The pronounced inter-layer correlation of $d_{3z^2-r^2}$ orbitals results in a downward trend and an extended flatness in the $\gamma$ band, mirroring the angle-resolved photoemission spectroscopy findings under ambient pressure.
Our numerical results provide important clues for understanding the strong correlation effects in La$_3$Ni$_2$O$_7$.
\end{abstract}

\date{\today}
\maketitle

\section{INTRODUCTION}
Recently, a new type of nickelate, La$_3$Ni$_2$O$_7$, has been discovered to exhibit high-temperature superconductivity under high pressures~\cite{sun2023nat}.
Subsequent theoretical study further characterized the electronic structure of La$_3$Ni$_2$O$_7$ using a bilayer two-orbital Hubbard model~\cite{luo2023prl}.
Up to now, several experimental works~\cite{hou2023cpl,zhang2024natphys,zhou2024arxiv,li2024arxiv,zhang2024effects,wang2024prx} have confirmed the zero resistance below $T_c$.
In addition, some theoretical studies~\cite{zhang2024natcomm,gu2023arxiv,liu2023prl,yang2023prbl,sakakibara2024prl,luo2024npjqm,zhang2023prb,tian2024prb,qin2023prbl,huang2023prb,jiang2024cpl,liu2023arxiv,fan2024prb,jiang2024prl,lu2024prl,qu2024prl} have revealed the pairing symmetry.
For the latest advancements on this material, one can refer to some review articles~\cite{wang2024cpl,yao2024kjdb}.
The basic structure of La$_3$Ni$_2$O$_7$ crystal consists of a Ni-O bilayer.
The electronic structure can be effectively described by a bilayer two-orbital model that includes Ni-$d_{x^2-y^2}$ and -$d_{3z^2-r^2}$ orbitals~\cite{sun2023nat,luo2023prl,gu2023arxiv}.
This bilayer two-orbital model has been widely used in theoretical study for La$_3$Ni$_2$O$_7$~\cite{zhang2024natcomm,gu2023arxiv,liu2023prl,liu2023prl,yang2023prbl,sakakibara2024prl,luo2024npjqm,zhang2023prb,tian2024prb,qin2023prbl,huang2023prb,jiang2024cpl,liu2023arxiv,zheng2023arxiv,wu2024scicpma,christiansson2023prl,shen2023cpl,lechermann2023prbl,yang2023prblI,PhysRevB.110.235119}.

In addition to being an important model for describing La$_3$Ni$_2$O$_7$, the bilayer two-orbital model, which is significantly different from the traditional single-orbital Hubbard model~\cite{anderson1987science,wen2006rmp,scalapino2012rmp}, serves as a new platform for investigating the ground-state phase diagram and interaction-induced correlation effects.
However, the determinant quantum Monte Carlo (DQMC) simulation on original bilayer two-orbital model encounters a sign problem, hindering the study of its ground-state properties at large system scales.
In this paper, we employ a simplified, half-filled bilayer two-orbital Hubbard model without sign problem to study correlation effects.
Although this simplified model may not fully capture the superconductivity in La$_3$Ni$_2$O$_7$, it is expected to reveal similar properties that could aid in understanding the electronic correlations within this material.
Furthermore, electron doping could potentially drive both orbitals in La$_3$Ni$_2$O$_7$ into half-filling case~\cite{luo2024npjqm}.

In this paper, using the projective formalism of DQMC method, we conduct large-scale simulations on a simplified bilayer two-orbital Hubbard model at half filling, revealing the ground-state phase diagram and dynamical properties.
We find an antiferromagnetic Mott insulator (AFMI) at $U>U_c$, a metallic phase in $U=0$, and a nonmagnetic weakly insulating (WI) phase for $0<U<U_c$.
Finite-size scaling analysis suggests that the magnetic transition between WI and AFMI is consistent with the three-dimensional (3D) classical Heisenberg universality class.
Additionally, we analyze the single-particle spectrum and dynamic spin structure factor, correlating our findings with experimental data from La$_3$Ni$_2$O$_7$, such as angle-resolved photoemission spectroscopy (ARPES).

\begin{figure}[t]
    \centering
    \includegraphics[width=\linewidth]{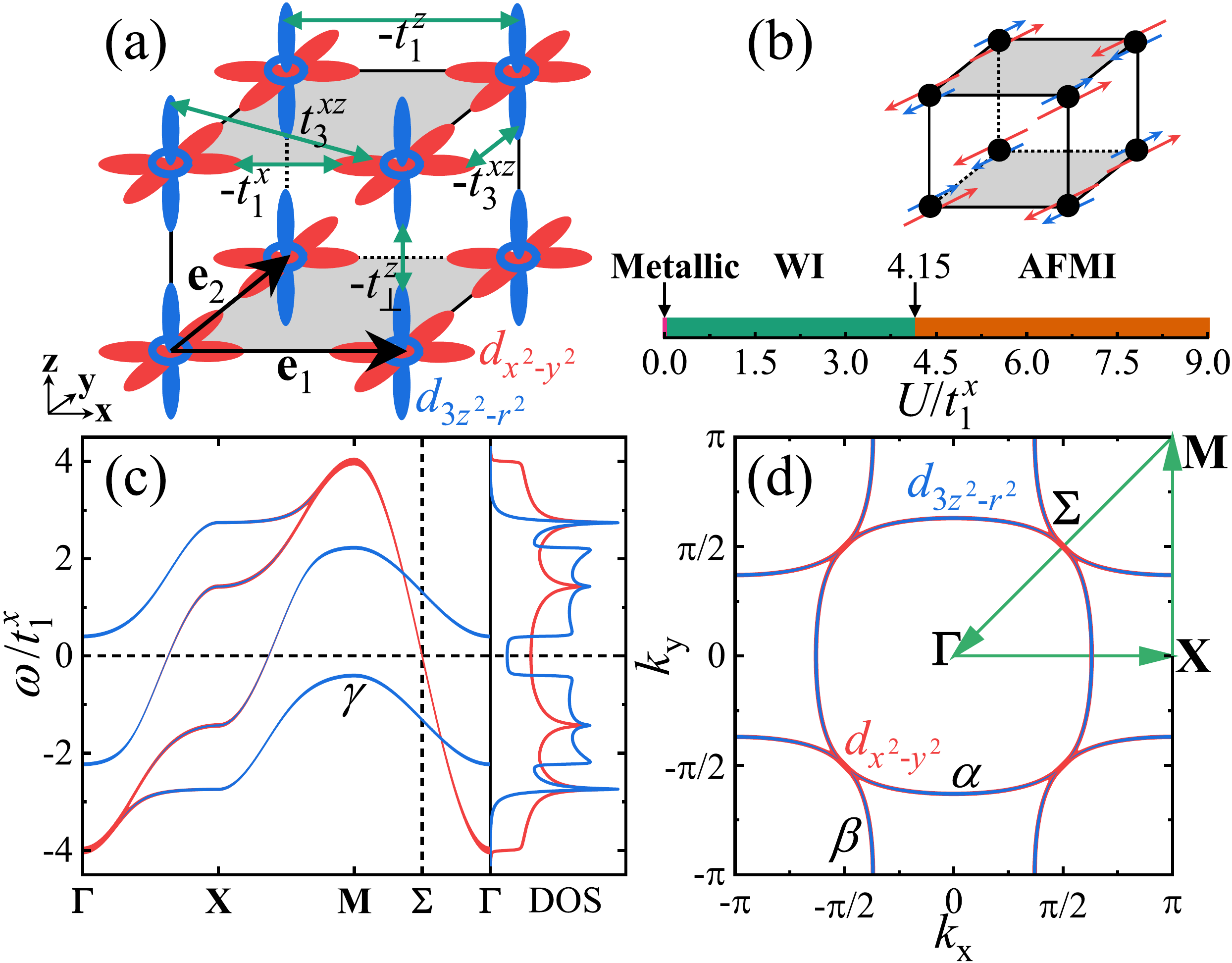}
    \caption{%
        (a)~Illustration of a simplified bilayer two-orbital model which contains $d_{x^2-y^2}$ (red) and $d_{3z^2-r^2}$ (blue) orbitals.
        (b)~Ground-state phase diagram.
        There are three phases: metallic phase at $U=0$, weakly insulating (WI) phase, and antiferromagnetic Mott insulator (AFMI).
        The configuration of AFMI is detailed with arrows indicating magnetic moments of two orbitals.
        (c)~Band structure and local density of states (DOS) in the noninteracting limit.
        The red and blue color represent the contributions from $d_{x^2-y^2}$ and $d_{3z^2-r^2}$ orbitals, respectively.
        The coordinates of the high-symmetry points are $\mathbf{\Gamma}=(0,0)$, $\mathbf{X}=(\pi,0)$, $\mathbf{M}=(\pi,\pi)$, and $\mathbf{\Sigma}=(\pi/2,\pi/2)$.
        (d)~Fermi surface at $U=0$.
        It is noted that only the $d_{x^2-y^2}$ orbital contributes to the density of states in $\mathbf{\Sigma}$ point at the Fermi level.}
    \label{fig:latt}
\end{figure}

\section{MODEL AND METHOD}
The Hamiltonian of bilayer two-orbital Hubbard model [Fig.~\ref{fig:latt}(a)] can be written as
\begin{eqnarray*}
H&=&\sum_{\langle i,j\rangle,\sigma}\sum_{m\alpha,n\lambda}t_{ij}^{m\alpha,n\lambda}c_{im\alpha\sigma}^{\dagger}c_{jn\lambda\sigma}\\
&&+U\sum_{i,m\alpha}(n_{im\alpha\uparrow}-\frac{1}{2})(n_{im\alpha\downarrow}-\frac{1}{2}),
\end{eqnarray*}
where $c_{im\alpha\sigma}^\dagger$ ($c_{im\alpha\sigma}$) creates (annihilates) an electron with spin-$\sigma$ in the $\alpha$ orbital of layer $m$ within the $i$-th unit cell, and $n_{im\alpha\sigma}=c_{im\alpha\sigma}^\dagger c_{im\alpha\sigma}$ is particle number operator.
$U>0$ is on-site Coulomb repulsive interaction which is set to the same value for each orbitals.
$t_{ij}^{m\alpha,n\lambda}$ represent hopping integrals [see Fig.~\ref{fig:latt}(a)], the values of which are set to be $t_1^x=1$ (as energy unit), $t_1^z=0.228$, $t_3^{xz}=0.495$, and $t_\perp^z=1.315$, ratios of which are inherited from Ref.~\cite{luo2023prl}.
But both the $d_{x^2-y^2}$ and $d_{3z^2-r^2}$ orbitals are at half filling in our case.

When $U=0$, the system is in a metallic phase with finite density of states at the Fermi level, as shown in Fig.~\ref{fig:latt}(c).
The Fermi surface, comprising an $\alpha$ pocket and a $\beta$ pocket that touch at the $\mathbf{\Sigma}$ point, is topologically equivalent to that of La$_3$Ni$_2$O$_7$ under ambient pressure~\cite{yang2024natcomm,jphu2024prb}.
In both cases, the $\gamma$ band, which is the lowest band near the $\mathbf{M}$ point and is contributed by $d_{3z^2-r^2}$ orbital, lies below the Fermi level.

When $U>0$, the projective formalism of determinant quantum Monte Carlo (DQMC) algorithm is employed to study the ground-state phases and their excitations, with details provided in Refs.~\cite{hirsch1985dqmc,assaad2008dqmc,assaad2022alf2.0}.
Due to the particle-hole symmetry and time-reversal symmetry, the DQMC simulations do not encounter the sign problem~\cite{wu2005prb}.
We use the projection imaginary-time lengths of $2\Theta\geqslant100$ to obtain the ground states and an imaginary-time step size of $\Delta\tau=0.05$ to reduce the system error of DQMC.
More details of the DQMC simulations can be found in Appendix~\ref{app:detail_dqmc}.
To reduce finite-size effects, our study uses two types of clusters with linear system sizes $L$ and $L^\prime$, corresponding to unit-cell numbers $N_c=L^2$ and $N_c=2L^{\prime2}$, respectively.
Additional information on these finite-size clusters is provided in Appendix~\ref{app:fs_clusters}.
In addition, unless otherwise noted, our simulations are conducted under periodic boundary conditions (PBC).

\begin{figure}[t]
    \centering
    \includegraphics[width=\linewidth]{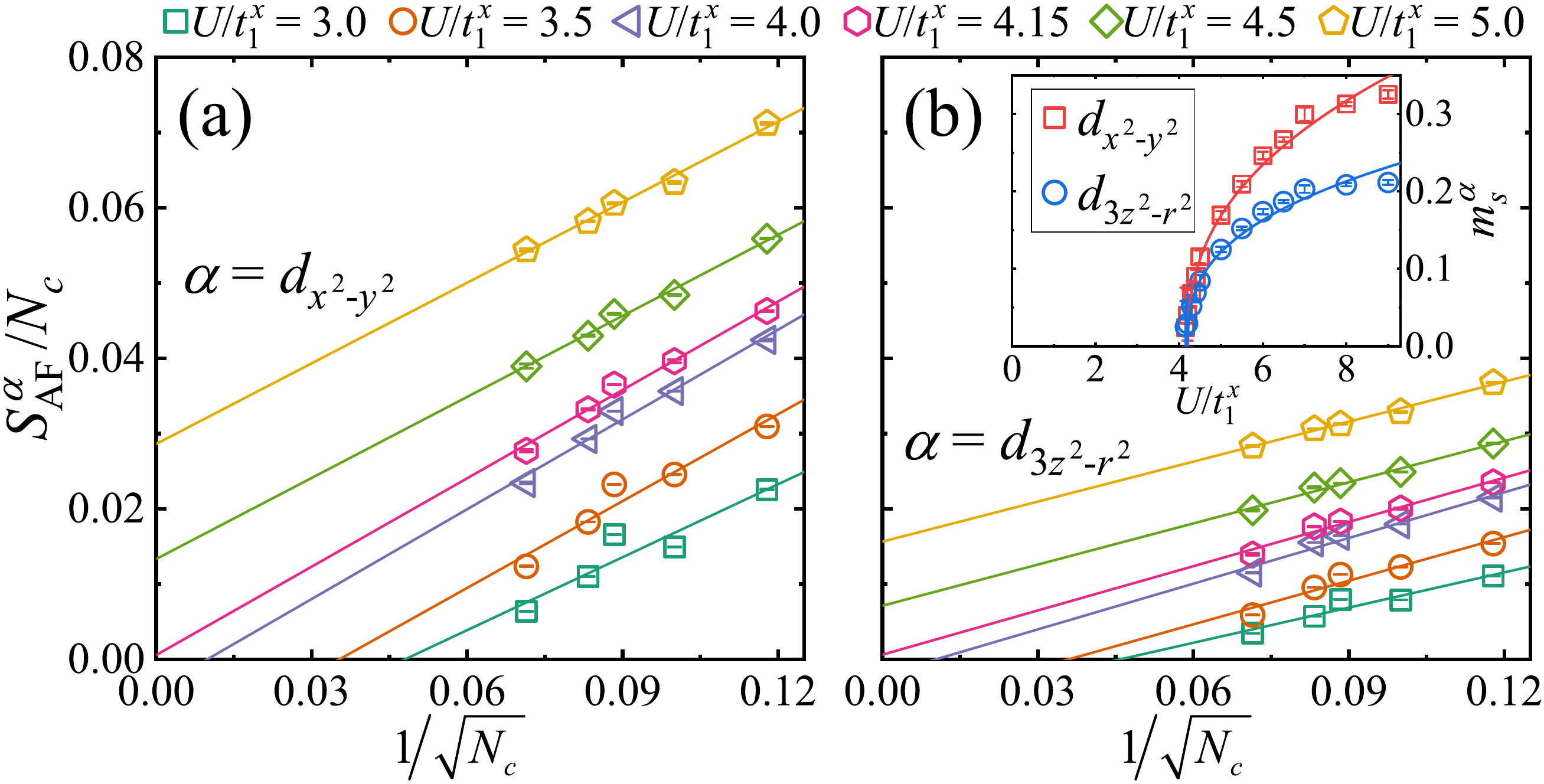}
    \caption{%
        Linear extrapolations of the spin structure factors $S_\text{AF}^\alpha/N_c$ for:
        (a)~the $d_{x^2-y^2}$ orbital and
        (b)~the $d_{3z^2-r^2}$ orbital.
        The linear cluster sizes used here are $L=10$, $12$, $14$ and $L^\prime=6$, $8$.
        Insert: The staggered magnetization are depicted as functions of $U$.
        The solid lines representing the fitting functions using $m_s^\alpha\propto(U-U_c^\prime)^\beta$, where $\beta$ is critical exponent of the order parameter.}
    \label{fig:safextra}
\end{figure}

\section{MAGNETIC TRANSITION}
The spin structure factor, written as $S^\alpha(\mathbf{k})=(1/2N_c)\sum_{i,j,m}\e^{-\i\mathbf{k}\cdot(\mathbf{R}_j-\mathbf{R}_i)}\langle\mathbf{S}_{im\alpha}\cdot\mathbf{S}_{jm\alpha}\rangle$, can be used to detect the magnetic ordering, where $\mathbf{S}_{im\alpha}=(1/2)\sum_{\sigma,\sigma^\prime}c_{im\alpha\sigma}^\dagger[\bm{\sigma}]_{\sigma\sigma^\prime}c_{im\alpha\sigma^\prime}$ is spin operator with Pauli matrix $\bm{\sigma}=(\sigma_x,\sigma_y,\sigma_z)$.
In the strong interaction region, an antiferromagnetic (AF) order is expected in the system, corresponding to a divergent tendency in spin structure factor at point $\mathbf{M}$, i.e., $S_\text{AF}^\alpha=S^\alpha(\mathbf{M})$.
And the staggered magnetization (between the same orbital) can be estimated as $m_s^\alpha=\lim\limits_{N_c\to\infty}\sqrt{S_\text{AF}^\alpha/N_c}$.
The linear extrapolations of $S_\text{AF}^\alpha/N_c$ are depicted in Fig.~\ref{fig:safextra}.
As the interaction strength increases to $U/t_1^x\approx4.15$, both the $d_{x^2-y^2}$ and $d_{3z^2-r^2}$ orbitals transition from nonmagnetic states to AF long-range orders, indicating a nearly identical magnetic transition point at $U_c/t_1^x=4.15(10)$ for both orbitals.
The inset of Fig.~\ref{fig:safextra}(b) illustrates the staggered magnetization as functions of $U$ for the different orbitals.
Figure~\ref{fig:latt}(b), upper panel, illustrates the AF order.
The increase in the inter-layer hopping $t_\perp^z$ within the $d_{3z^2-r^2}$ orbitals and the hybridization hopping $t_3^{xz}$ between the $d_{x^2-y^2}$ and $d_{3z^2-r^2}$ orbitals both suppresses the AF order.
For a specific discussion on the effects of varying $t_\perp^z$ and $t_3^{xz}$ on the magnetic correlation, see Appendix~\ref{app:effects_other_hop_param}.

\section{WEAKLY INTERACTING REGION}
\begin{figure}[t]
    \centering
    \includegraphics[width=\linewidth]{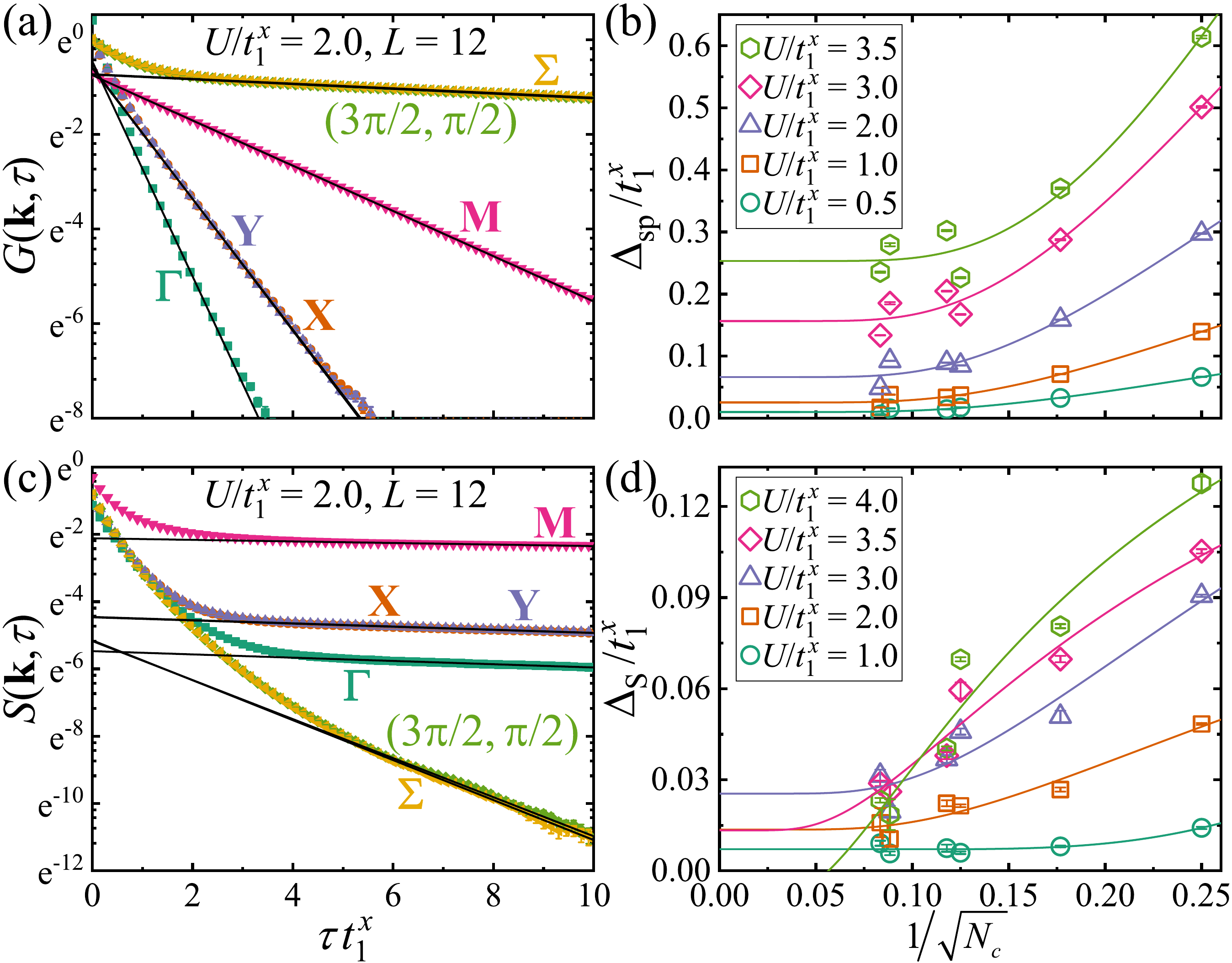}
    \caption{%
        (a)~Imaginary time-displaced single-particle Green's function $G(\mathbf{k},\tau)$ and
        (c)~imaginary time-displaced spin correlation function $S(\mathbf{k},\tau)$ at high-symmetry points are plotted on a semilog scale for $U/t_1^x=2.0$ and $L=12$, where $\mathrm{Y}=(0,\pi)$.
        The black solid lines indicate linear fits to the tails to extract the corresponding gaps.
        Finite-size extrapolations of the
        (b)~single-particle gap $\Delta_\text{sp}$ and
        (d)~spin gap $\Delta_\text{S}$, where the fitting function takes the form $f(x)=a+b\e^{-c/x}$~\cite{yang2022prb,hohenadler2012prb}.}
    \label{fig:gapextra}
\end{figure}

After identifying the magnetic transition at finite $U$, two critical questions emerge: First, whether the magnetic transition takes place between two insulating phases or coincides with the metal-insulator transition.
Second, whether there is a single phase or multiple phases present in the weakly interacting region where $0<U<U_c$.
To address the first question, we calculate the single-particle gap $\Delta_\text{sp}$, which can be derived from the imaginary time-displaced single-particle Green's function $G(\mathbf{k},\tau)=(1/4N_c)\sum_{i,j,m\alpha,\sigma}\e^{-\i\mathbf{k}\cdot(\mathbf{R}_j-\mathbf{R}_i)}\langle c_{im\alpha\sigma}^\dagger(\tau)c_{jm\alpha\sigma}(0) \rangle$, here $c_{im\alpha\sigma}^\dagger(\tau)=\e^{\tau H}c_{im\alpha\sigma}^\dagger\e^{-\tau H}$, and $\tau$ is the imaginary time.
When $\tau$ is sufficiently large, the Green's function behaves as $G(\mathbf{k},\tau)\propto\e^{-\tau\Delta_\text{sp}(\mathbf{k})}$.
As depicted in Fig.~\ref{fig:gapextra}(a), the smallest gap is observed at $\mathbf{\Sigma}$ Fermi wave vector in finite-size system.
The finite-size extrapolations of $\Delta_\text{sp}$ are presented in Fig.~\ref{fig:gapextra}(b).
The fitting function used is $f(x)=a+b\e^{-c/x}$~\cite{yang2022prb,hohenadler2012prb}.
It is estimated that the single-particle gap is already finite at the $\mathbf{\Sigma}$ point when $U/t_1^x\gtrsim0.5$.
To ascertain whether the single-particle gap is open across the entire Fermi surface simultaneously, we selected an $L^\prime=8$ cluster that encompasses several Fermi wave vectors within its Brillouin zone.
The finite-size single-particle gaps at these wave vectors are nearly identical in the $0<U<U_c$ region (refer to Appendix~\ref{app:simult_spgap}).
This observation suggests that the gap at the $\mathbf{\Sigma}$ point is indicative of the system's single-particle gap.
Consequently, we can reasonably conclude that the system develops a full single-particle gap across the entire Fermi surface, thereby entering a weakly insulating phase.

Similarly, the spin gap $\Delta_\text{S}$ can be extracted from the imaginary time-displaced spin correlation function $S(\mathbf{k},\tau)=(1/4N_c)\sum_{i,j,m\alpha}\e^{-\i\mathbf{k}\cdot(\mathbf{R}_j-\mathbf{R}_i)}\langle\mathbf{S}_{im\alpha}(\tau)\cdot\mathbf{S}_{jm\alpha}(0)\rangle$.
As shown in Fig.~\ref{fig:gapextra}(c), the smallest spin gap is observed at the $\mathbf{M}$ point, which is associated with the wave vector of the AF order.
Figure~\ref{fig:gapextra}(d) illustrates the finite-size extrapolations of the spin gap.
As $U$ increases, the spin gap increases with $U$, reaching a maximum ($U/t_1^x\sim3.0$) before diminishing to zero in the AF phase.
The extrapolated negative spin gap at $U/t_1^x=4.0$, which is less than $U_c/t_1^x$, may be attributed to the finite-size effects.
In summary, Figs.~\ref{fig:gapextra}(b) and \ref{fig:gapextra}(d) support a gapped phase in the weakly interacting region and the existence of a finite $U$ magnetic transition between two insulating phases.

\begin{figure}[t]
    \centering
    \includegraphics[width=\linewidth]{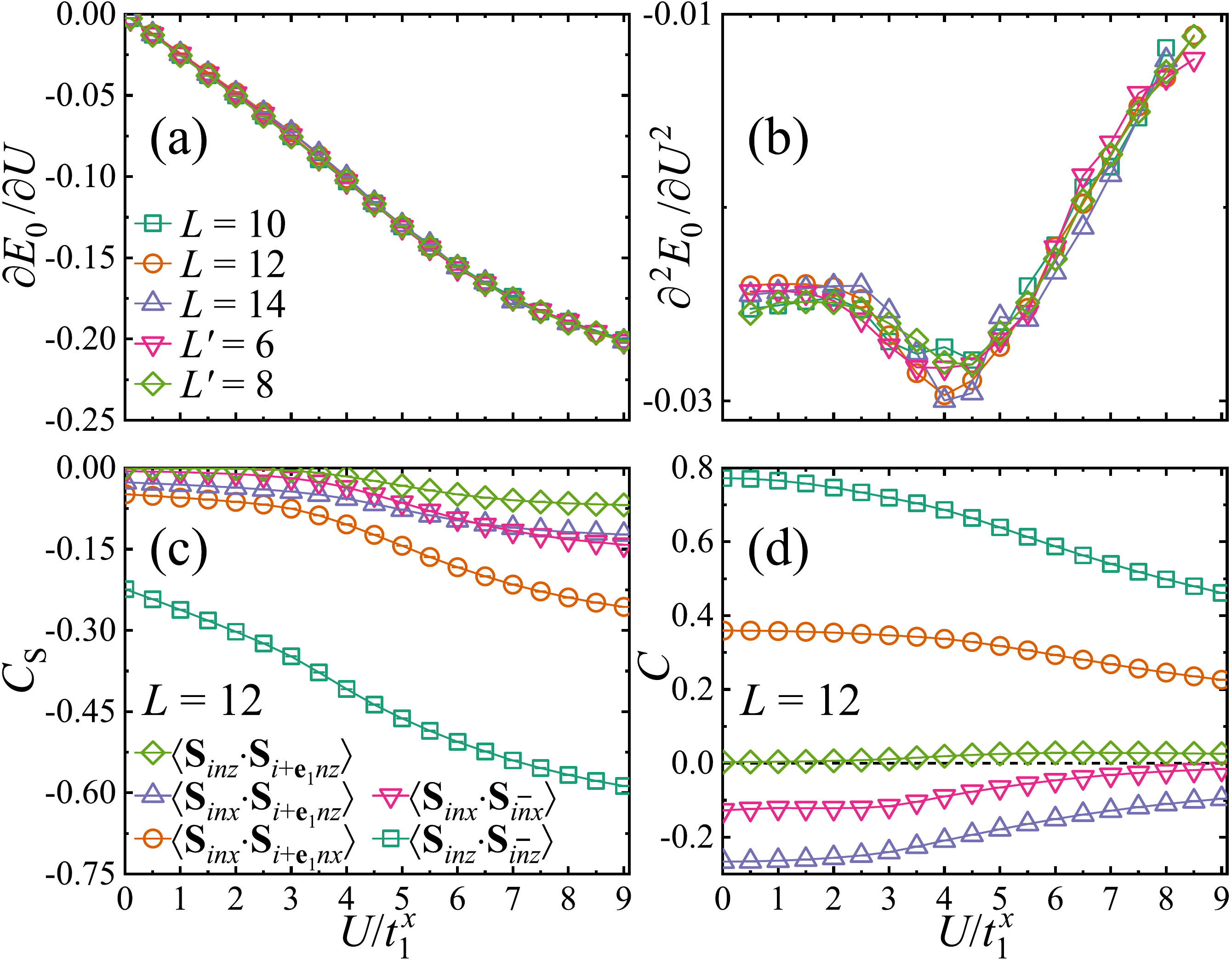}
    \caption{%
        (a)~The first derivative of energy with respect to $U$ and
        (b)~the second derivative of energy with respect to $U$ as functions of $U$ for different cluster sizes.
        (c)~Local spin correlations and
        (d)~local single-particle correlations as functions of $U$ for a cluster size of $L=12$.
        In panel (d), the symbols correspond to the same bonds as in (c).
        The subscripts $n$ and $\bar{n}$ indicate opposite layers, while $x$ and $z$ represent the $d_{x^2-y^2}$ and $d_{3z^2-r^2}$ orbitals, respectively.}
    \label{fig:wichara}
\end{figure}

To address the second question regarding the absence of a quantum phase transition within the range $0<U<U_c$, we show the first derivative of the ground-state energy $E_0$, which is defined as: $\partial E_0/\partial U=(1/4N_c)\sum_{i,m\alpha}\langle(n_{im\alpha\uparrow}-1/2)(n_{im\alpha\downarrow}-1/2)\rangle$, where $E_0=\langle H\rangle/(4N_c)$.
The discontinuity in $\partial E_0/\partial U$ can be indicative of phase transitions~\cite{kancharla2007prb,tahara2008locmom,sentef2009prb,euverte2013prb,wang2020prb}.
Figure~\ref{fig:wichara}(a) shows $\partial E_0/\partial U$ decreasing monotonically with $U$ for various cluster sizes.
Furthermore, the numerical first derivative of $\partial E_0/\partial U$, which corresponds to the second derivative of energy [as shown in Fig.~\ref{fig:wichara}(b)], exhibits only single peak around the magnetic transition point $U_c$.
This observation implies that there is no additional quantum phase transition occurring within the interval $0<U<U_c$.

To further characterize the weakly insulating phase, we have extrapolated various conventional order parameters, including charge-density wave, pairing, interlayer exciton, and bond-order wave.
However, as detailed in Appendix~\ref{app:absence_conven_order}, all these orders extrapolate to zero, suggesting that the system enters a featureless insulating phase with dominating local correlation.
In this ground state, the $d_{3z^2-r^2}$ orbitals tend to form a direct product of the ground state of a two-site Hubbard model.
This tendency is due to the strong hopping ratios $t_\perp^z/t_1^z=5.768$ and $t_\perp^z/t_3^{xz}=2.657$.
Additionally, the hybridization $t_3^{xz}$ between nearby $d_{3z^2-r^2}$ and $d_{x^2-y^2}$ orbitals allows the $d_{x^2-y^2}$ orbital to generate a mass gap, contributing to the insulating phase.
This weakly insulating phase may belong to a symmetric mass generation (SMG) phase without symmetry breaking~\cite{ayyar2015prd,he2016prb,wang2022sym,lu2023arxiv,chang2023arxiv}.
Figures~\ref{fig:wichara}(c) and \ref{fig:wichara}(d) display the local real-space spin correlation $C_\text{S}=\langle\mathbf{S}_{im\alpha}\cdot\mathbf{S}_{jn\lambda}\rangle$ and the single-particle correlation $C=(1/2)\sum_\sigma\langle c_{im\alpha\sigma}^\dagger c_{jn\lambda\sigma}+H.c.\rangle$ as functions of $U$, respectively.
As expected, the interlayer $d_{3z^2-r^2}$ orbitals exhibit the strongest spin correlation $\langle\mathbf{S}_{inz}\cdot\mathbf{S}_{i\bar{n}z}\rangle$ and single-particle correlation $(1/2)\sum_\sigma\langle c_{inz\sigma}^\dagger c_{i\bar{n}z\sigma}+H.c.\rangle$.

\begin{figure}[t]
	\centering
	\includegraphics[width=\linewidth]{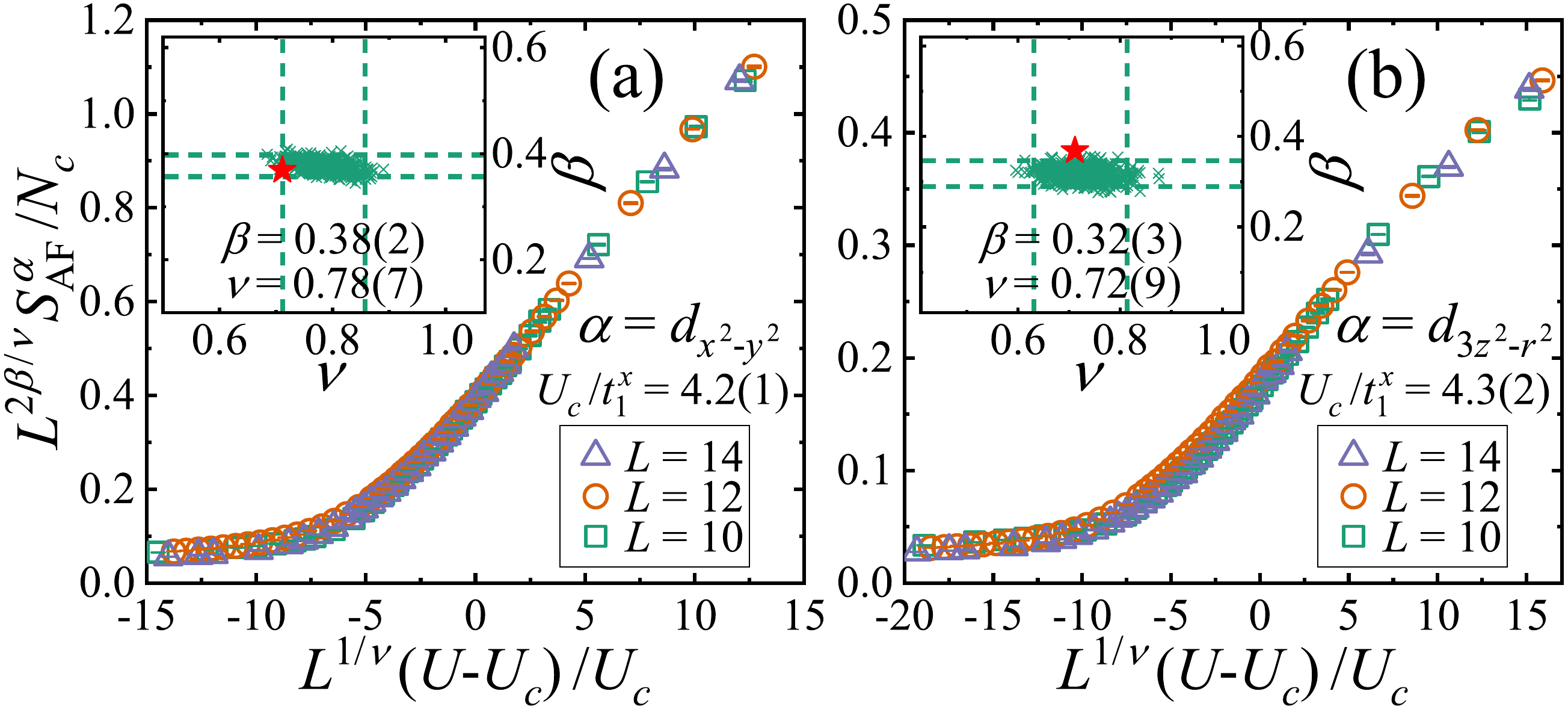}
	\caption{%
            Finite-size scaling of the spin structure factor for the
            (a)~$d_{x^2-y^2}$ and
            (b)~$d_{3z^2-r^2}$ orbitals, respectively.
            Insets show the scatter plots derived from Monte Carlo samplings, with horizontal and vertical dashed lines indicating the 95\% confidence limits.
            The red stars denote the 3D O(3) critical exponents~\cite{campostrini2002prb}.}
	\label{fig:fssa}
\end{figure}

\section{CRITICAL BEHAVIOR}
From the perspective of symmetry analysis, the transition from WI phase, featureless with charge and spin gaps, to the AFMI phase, which breaks the SU(2) spin-rotation symmetry, is classified within the 3D O(3) universality class.
By employing the standard scaling analysis using formula $S_\text{AF}^\alpha/N_c=L^{-2\beta/\nu}f[L^{1/\nu}(U-U_c)/U_c]$, we numerically investigate the critical behavior of the WI-AFMI transition.
Here, $\beta$ and $\nu$ are critical exponents.
The optimal critical parameters are determined through Bayesian scaling analysis, which relies on Gaussian process regression, as detailed in Refs.~\cite{harada2011pre,harada2015pre}.
The outcomes of the data collapse are depicted in Fig.~\ref{fig:fssa}, with confidence interval estimations derived from Monte Carlo sampling.
The quantum critical point identified here is in close proximity to the value obtained from the extrapolations of the spin structure factor $U_c/t_1^x\approx4.15$, thereby reinforcing the existence of a finite $U$ magnetic transition.
The critical exponents obtained are nearly identical to those of the 3D O(3) Heisenberg universality class~\cite{campostrini2002prb} with minor deviations.
The deviations in the critical exponents may be due to the limitation of finite-size lattices~\cite{jia2024prb}.

\begin{figure}[t]
	\centering
	\includegraphics[width=\linewidth]{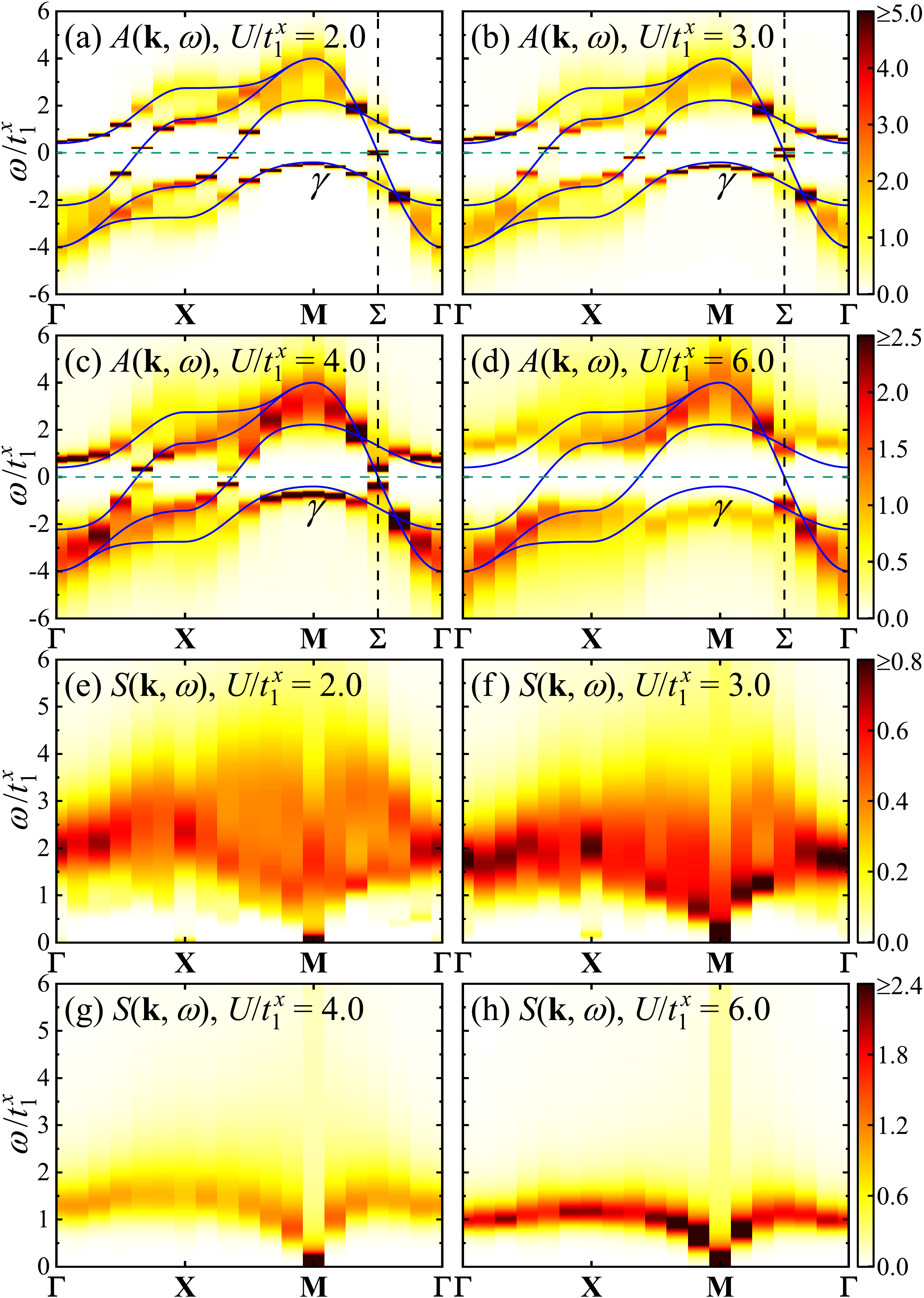}
	\caption{%
            (a)-(d)~Single-particle spectra and
            (e)-(h)~dynamic spin structure factors along high-symmetry path for various values of $U$.
            The blue solid lines in panels (a)-(d) represent the noninteracting band structures.}
	\label{fig:dynspe}
\end{figure}

\section{DYNAMIC EXCITATION SPECTRA}
The single-particle spectral function $A(\mathbf{k},\omega)$ and the dynamic spin structure factor $S(\mathbf{k},\omega)$, respectively, reflect charge and spin excitations, which can be directly measured experimentally.
Numerically, the dynamic excitation spectra can be computed using the stochastic maximum entropy algorithm~\cite{assaad2022alf2.0,sandvik1998prb,beach2004arxiv,shao2023physrep} based on $G(\mathbf{k},\tau)=(1/\pi)\int\d\omega\,A(\mathbf{k},\omega)\e^{-\tau\omega}$ and $S(\mathbf{k},\tau)=(1/\pi)\int\d\omega\,S(\mathbf{k},\omega)\e^{-\tau\omega}$.
The numerical results for the single-particle spectral functions $A(\mathbf{k},\omega)$ with $L=12$ for various values of $U$ are depicted in Figs.~\ref{fig:dynspe}(a)-(d).
When $U$ is small, such as $U/t_1^x=2.0$, the band dispersion can still be accurately described by the noninteracting limit.
Moreover, a small single-particle gap is observed at the $\mathbf{\Sigma}$ point (see Appendix~\ref{app:more_sme_reslts}).
As $U$ increases, the single-particle gap widens, and the bands become blurred due to interaction-induced correlation effects, rendering quasiparticles ill defined.
Most notably, the $\gamma$ band, which is the lower Hubbard band near the Fermi level around the $\mathbf{M}$ point and predominantly contributed by the $d_{3z^2-r^2}$ orbital, progressively moves away from the Fermi level as $U$ increases.
The deviation of the $\gamma$ band at the $\mathbf{M}$ point from the noninteracting case follows the same trend as the growth of the single-particle gap at the $\mathbf{\Sigma}$ point, indicative of a Mottness behavior by shifting lower Hubbard band as $U$ increases.
The flatness of the $\gamma$ band is attributed to the strong localization of the $d_{3z^2-r^2}$ orbitals.

The dynamic spin structure factors $S(\mathbf{k},\omega)=(1/4)\sum_{m\alpha}S^{m\alpha}(\mathbf{k},\omega)$ are shown in Figs.~\ref{fig:dynspe}(e)-(h).
In WI phase regime, the spin gaps are very small [on the order $O(10^{-2})t_1^x$] at the $\mathbf{\Gamma}$, $\mathbf{X}$, and $\mathbf{M}$ points, making them challenging to discern in the contour plots shown in Figs.~\ref{fig:dynspe}(e) and \ref{fig:dynspe}(f).
Additionally, the low-energy spectral weights at $\mathbf{\Gamma}$ and $\mathbf{X}$ are significantly weaker compared to those at the $\mathbf{M}$ point.
At higher energy, a broad continuum is observed across a substantial region of the $\mathbf{k}$-$\omega$ plane, characterized by weak intensity due to ``itinerant" weakly correlated electrons.
As $U$ increases, a divergent spectral weight develops around $\omega/t_1^x=0$ at the $\mathbf{M}$ point, signaling the emergence of antiferromagnetic order.
Concurrently, the low-energy spectral weights at $\mathbf{X}$ and $\mathbf{\Gamma}$ diminish, leading to the emergence of well-defined magnon spectra as the dominant features.

\section{CONCLUSION AND DISCUSSION}
Through the large-scale DQMC simulations, we have determined the ground-state phase diagram, critical behavior, and dynamical excitation of a simplified bilayer two-orbital Hubbard model at half filling.
Our findings reveal an intermediate weakly insulating phase between the metallic phase at $U/t_1^x=0$ and the AFMI at $U/t_1^x>4.15(10)$.
This intermediate phase with weak charge and spin gaps may represent a symmetric mass-generation phase with significant interlayer correlations between the $d_{3z^2-r^2}$ orbitals.
The quantum phase transition from the WI phase to the AFMI is consistent with the 3D classical Heisenberg universality class.

Although our simplified model may deviate from the La$_3$Ni$_2$O$_7$ under ambient pressure, it shares a similar Fermi surface and exhibits common strong correlation features.
For example, ARPES experiments have observed a flat $d_{3z^2-r^2}$ bonding band below the Fermi level, indicative of strong electron correlation~\cite{yang2024natcomm}, which is quite similar to our findings in single-particle spectral function.
It is also noteworthy that recent experimental measurements have indicated a spin-density-wave instability with a wave vector $(\pi/2,\pi/2)$ in La$_3$Ni$_2$O$_7$~\cite{wang2024cpl,chen2024nc,dan2024arxiv,khasanov2024arxiv,chen2024prl}, contrasting with the magnetic ordering observed in our study.
This discrepancy could stem from differences in electron filling of the $d_{x^2-y^2}$ orbital and doubling of unit cell at low pressure.
Reconsidering some neglected terms in the simplified model to do the DQMC calculations could be an intriguing avenue for future research.
Addressing the sign problem in DQMC to extract useful information and gain a
deeper understanding of the density wave in La$_3$Ni$_2$O$_7$ under ambient pressure remains a challenge.

\begin{acknowledgments}
We would like to thank Yi-Zhuang You, Wei Zhu, W\'ei W\'u, and Zhihui Luo for helpful discussions. This project is supported by NKRDPC-2022YFA1402802, NSFC-12474248, NSFC-92165204, NSFC-12494590, Leading Talent Program of Guangdong Special Projects (Grant No. 201626003), Guangdong Basic and Applied Basic Research Foundation (Grant No. 2023B1515120013), Youth S\&T Talent Support Programme of Guangdong Provincial Association for Science and Technology (GDSTA) (Grant No. SKXRC202404), Research Center for Magnetoelectric Physics of Guangdong Province (2024B0303390001), and Guangdong Provincial Quantum Science Strategic Initiative (GDZX2401010).
\end{acknowledgments}

\appendix

\section{\label{app:detail_dqmc}DETAILS OF DQMC SIMULATIONS}
We employ the projective formalism of DQMC~\cite{hirsch1985dqmc,assaad2008dqmc,assaad2022alf2.0} for large-scale numerical simulations, which obtains the ground state by projecting a trial wave function:
\begin{eqnarray*}
\frac{\langle\Psi_0|O|\Psi_0\rangle}{\langle\Psi_0|\Psi_0\rangle}=\lim_{\Theta\to+\infty}\frac{\langle\Psi_T|\e^{-\Theta H}O\e^{-\Theta H}|\Psi_T\rangle}{\langle\Psi_T|\e^{-2\Theta H}|\Psi_T\rangle},
\end{eqnarray*}
where $O$ is an operator, $|\Psi_0\rangle$ is the many-body ground state, $|\Psi_T\rangle$ is the trial wave function, and $\Theta$ is the projection parameter.
We use an asymmetric Suzuki-Trotter decomposition for imaginary-time discretization and decouple the interaction term using a four-component Hubbard-Stratonovich transformation.
In the simulations, we typically use the projection lengths of $2\Theta\geqslant100$ to obtain the ground states and fix the imaginary-time discretization $\Delta\tau=0.05$ to reduce the system error $\mathcal{O}(\Delta\tau^2)$.
For each Markov chain, we treat the first bin as the thermalization process, typically with each bin containing 50 sweeps ($\sim$$10^7$ Monte Carlo steps).
When measuring the equal-time correlations, we perform measurements at each $\tau$ value within a range of $2\Theta/5$ in the middle of $2\Theta$.
Non-equal-time correlations are measured during sweep up at 200 imaginary-time $\tau$ values in the middle of $2\Theta$.
We typically calculate the number of bins with $N_b\geqslant10$.
For measurements of the non-equal-time correlations that required analytic continuation, we set $N_b\geqslant20$ to obtain reliable estimates of the covariance matrix.
We perform the measurements using parallelization, and run eight independent Markov chains.
The sign problem in DQMC can be avoided through a unitary particle-hole transformation
\begin{eqnarray*}
c_{im\alpha\downarrow}^\dagger&\to&(-1)^{i+m}c_{im\alpha\downarrow},\\
c_{im\alpha\downarrow}&\to&(-1)^{i+m}c_{im\alpha\downarrow}^\dagger,
\end{eqnarray*}
where $m=1$, $2$ denotes the layer index.
Due to the particle-hole symmetry, this transformation ensures that the weight of any DQMC auxiliary-field configuration is a non-negative real number~\cite{wu2005prb}.

\section{\label{app:fs_clusters}FINITE-SIZE CLUSTERS}
The bilayer two-orbital Hubbard model we investigated belongs to a square lattice, whose primitive vectors are given by $\mathbf{e}_1=(1,0)a$ and $\mathbf{e}_2=(0,1)a$, where $a\equiv1$ represents the lattice constant, serving as the unit of length.
In the main text, to mitigate finite-size effects, our study employs two types of finite-size clusters.
The first type, expanded by primitive vectors $\mathbf{e}_1$ and $\mathbf{e}_2$, has a linear cluster size denoted by $L$, with the number of unit cells $N_c=L^2$, reaching up to $L=14$.
The second type, expanded by lattice vectors $\mathbf{a}_1=(1,-1)a$ and $\mathbf{a}_2=(1,1)a$, has a linear cluster size denoted by $L^\prime$, with $N_c=2L^{\prime2}$ unit cells, up to $L^\prime=8$.
As discussed in the next paragraph, the selection of these lattices is based on their ability to accommodate the $\mathbf{M}$ and $\mathbf{\Sigma}$ wave vectors, crucial for detecting magnetic order and insulating behavior under Hubbard $U$ interaction.

\begin{figure}[t]
	\centering
	\includegraphics[width=\linewidth]{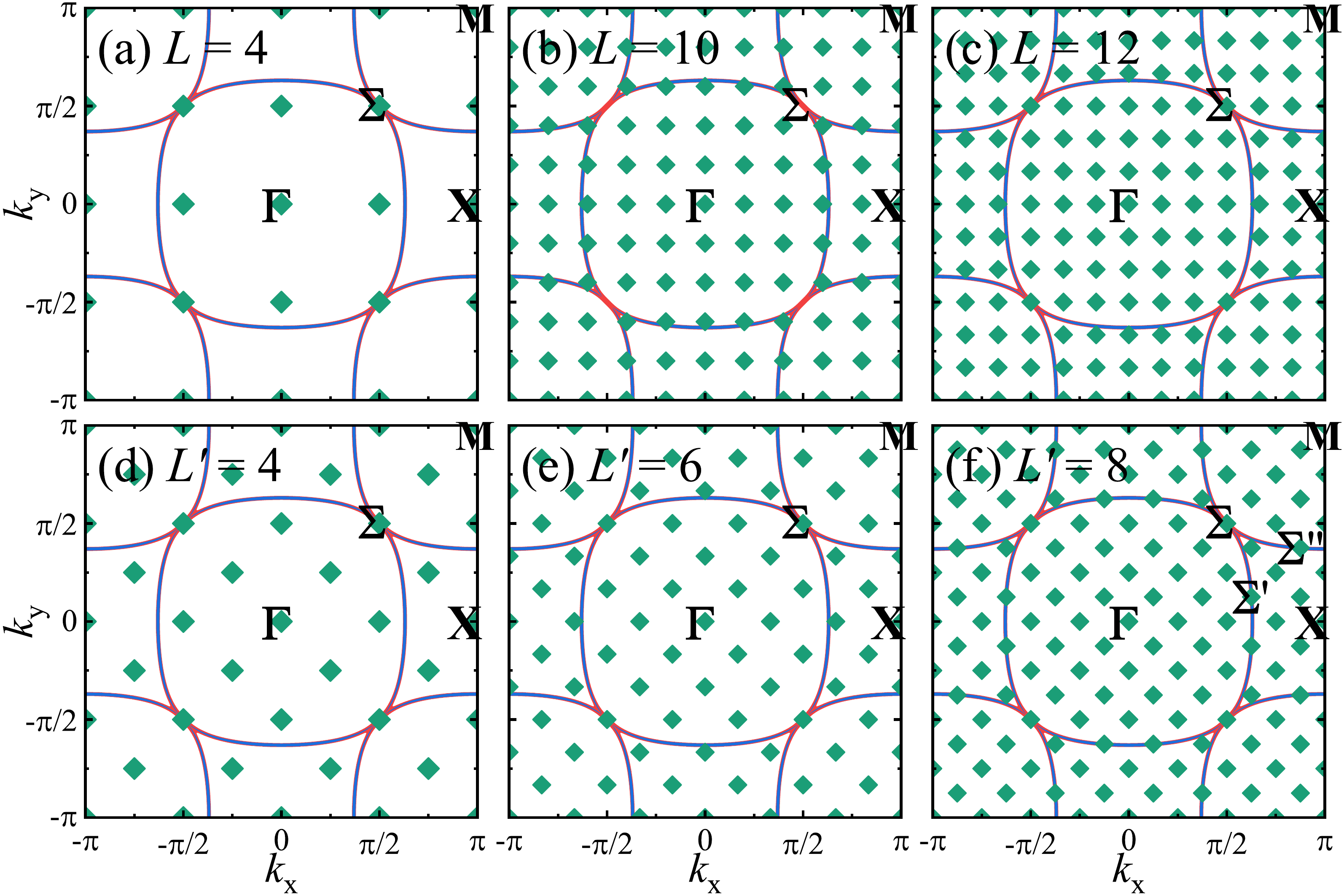}
	\caption{%
		Panels~(a)-(f) display the wave-vector points (represented by green squares) for different finite-size clusters.
		In (f), $\mathbf{\Sigma}^\prime=(5\pi/8,\pi/8)$ and $\mathbf{\Sigma}^{\prime\prime}=(7\pi/8,3\pi/8)$.
		For comparison, the Fermi surface is outlined with solid lines.
		The Fermi surface is derived from the spectral function $A(\mathbf{k},\omega)$ at $\omega=0$ in noninteracting limit $U=0$.
		The red and blue color represent the contributions from $d_{x^2-y^2}$ and $d_{3z^2-r^2}$ orbitals, respectively.
		It is noted that the $\mathbf{\Sigma}$ point only has contribution from the $d_{x^2-y^2}$ orbital.}
	\label{fig:kspace}
\end{figure}

In the weakly interacting region, the Fermi surface plays a pivotal role in determining the physical properties of the system.
Small-sized clusters may exhibit behaviors that are significantly different from those of larger clusters because their finite-size wave vectors are located away from the Fermi surface in the Brillouin zone.
In Fig.~\ref{fig:kspace}, we illustrate a selection of finite-size wave vectors for various system sizes and compare them with the Fermi surface.
The $\mathbf{M}=(\pi,\pi)$ point corresponds to the magnetic Bragg peak position for a large on-site Hubbard $U$.
We consistently choose clusters under PBC that include the $\mathbf{M}$ point within the Brillouin zone.
In the main text, the clusters used for extrapolating the antiferromagnetic structure factor are those with $L\geqslant10$ and $L^\prime\geqslant6$ where the wave-vector points are sufficiently dense, and the finite-size effects are minimal.
Linear fittings provide reliable extrapolated values for these clusters.
To probe the single-particle gap at the Fermi surface, we not only consider the $\mathbf{M}$ point but also require that the finite-sized clusters we select encompass the $\mathbf{\Sigma}=(\pi/2,\pi/2)$ point within the Brillouin zone.
Consequently, we use six finite-size clusters that include these two wave vectors, specifically $L=4$, $8$, $12$ and $L^\prime=4$, $6$, $8$, for extrapolating the gaps.

\begin{figure}[t]
	\centering
	\includegraphics[width=\linewidth]{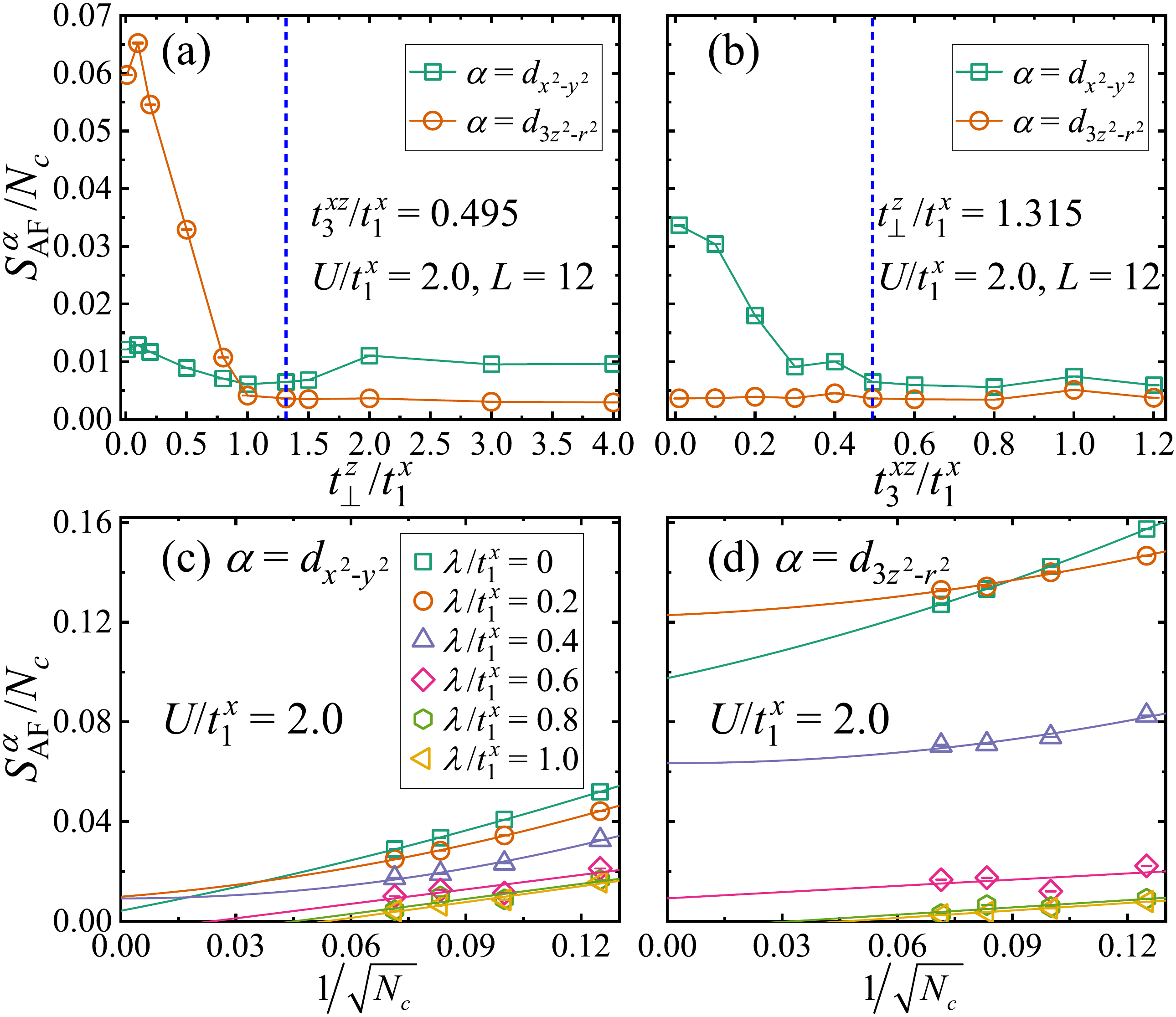}
	\caption{%
		Static spin-structure factors at the $\mathbf{M}$ point as functions of
		(a)~$t_\perp^z$ and
		(b)~$t_3^{xz}$, with $U/t_1^x=2.0$ and $L=12$ held constant.
		In each case, the value of other variable is kept fixed while one is varied.
		Extrapolation of the spin-structure factor under $U/t_1^x=2.0$ and different $\lambda$ for the
		(c)~$d_{x^2-y^2}$ and
		(d)~$d_{3z^2-r^2}$ orbitals, respectively.
		The data for $\lambda/t_1^x=0$, $0.2$, and $0.4$ are fitted using a second-order polynomial, while for the others, linear fitting is used due to the data oscillations.}
	\label{fig:tpt3eff}
\end{figure}

\section{\label{app:effects_other_hop_param}SUPPRESSING MAGNETIC CORRELATION BY INTERLAYER HOPPING $t_\perp^z$ AND HYBRIDIZATION HOPPING $t_3^{xz}$}
To further investigate the effects of interlayer hopping $t_\perp^z$ and the hybridization hopping $t_3^{xz}$, we calculate the static spin-structure factors at the $\mathbf{M}$ point as functions of $t_\perp^z$ while keeping $t_3^{xz}/t_1^x=0.495$ constant, as shown in Fig.~\ref{fig:tpt3eff}(a).
It is evident that an increase in $t_\perp^z$ significantly suppresses the magnetic correlation in the $d_{3z^2-r^2}$ orbital.
When varying $t_3^{xz}$ with $t_\perp^z/t_1^x$ held constant at $1.315$, the spin correlation in $d_{x^2-y^2}$ orbital is observed to be suppressed.
In the case of the $d_{3z^2-r^2}$ orbital, the spin-structure factor remains minimal and nearly unchanged with $t_\perp^z$.
This stability may be attributed to the strong local interlayer correlations induced by $U$ and the substantial $t_\perp^z$.

In order to investigate the combined effects of $t_\perp^z$ and $t_3^{xz}$ on the magnetic ordering of the system, we then set $t_\perp^z=1.315\lambda$ and $t_3^{xz}=0.495\lambda$, and perform DQMC simulations for different value of $\lambda$ at $U/t_1^x=2.0$.
The finite-size extrapolations of the spin-structure factor are depicted in Figs.~\ref{fig:tpt3eff}(c) and \ref{fig:tpt3eff}(d).
When $\lambda/t_1^x=0$, the system decouples into four independent single-layer single-orbital models on square lattices, each exhibiting an antiferromagnetic long-range order.
When $\lambda/t_1^x$ increases beyond $0.6$, the antiferromagnetic order disappears.
Therefore, during the process of increasing $\lambda/t_1^x$ from zero to the value in the main text ($\lambda/t_1^x=1.0$), there is indeed a transition from an antiferromagnetic ordered phase to a nonmagnetic phase, although due to significant finite-size effects, it is difficult to determine the exact transition point $\lambda_c$.

\begin{figure}[t]
	\centering
	\includegraphics[width=0.8\linewidth]{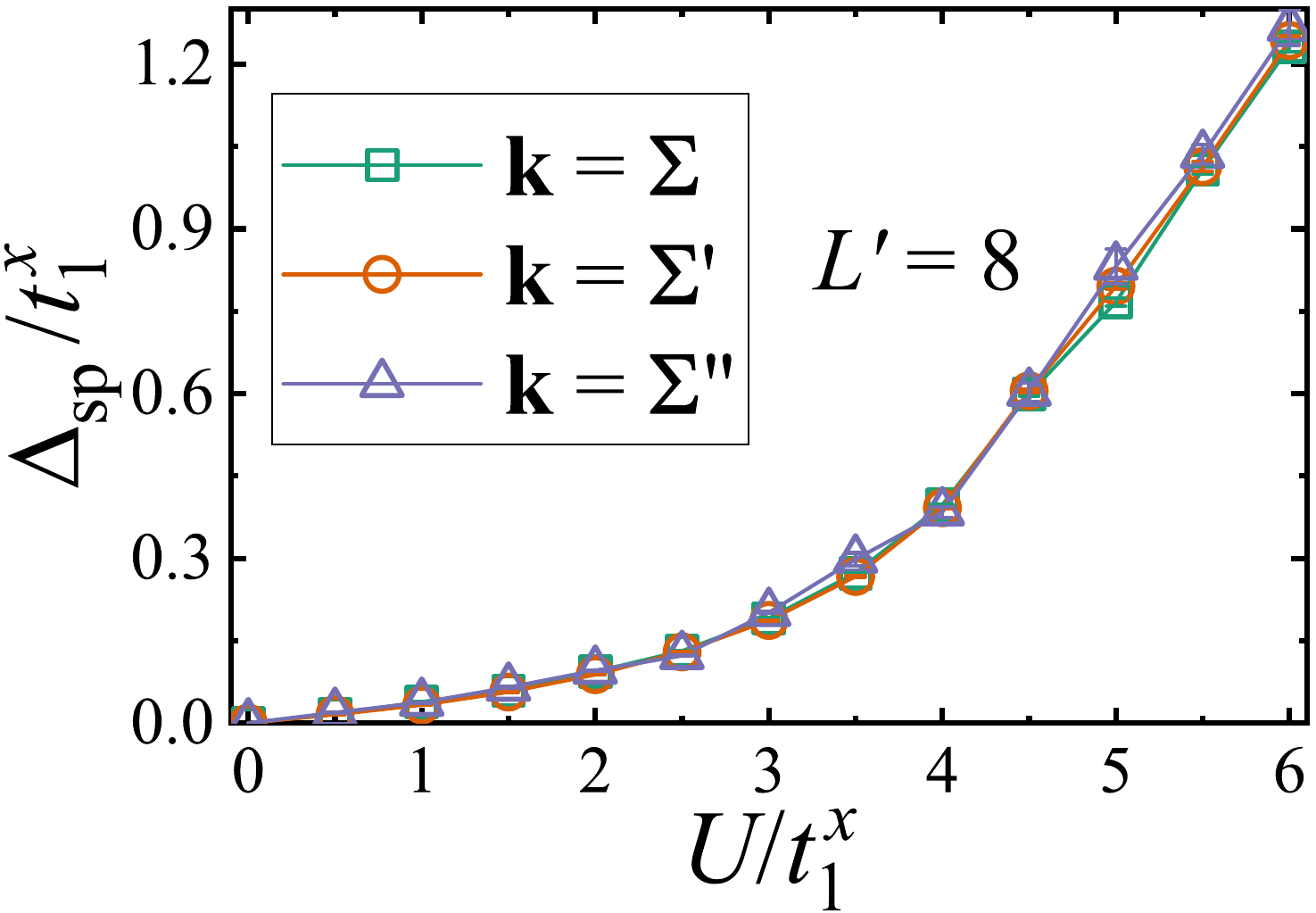}
	\caption{%
		The single-particle gaps at the $\mathbf{\Sigma}=(\pi/2,\pi/2)$, $\mathbf{\Sigma}^\prime=(5\pi/8,\pi/8)$, and $\mathbf{\Sigma}^{\prime\prime}=(7\pi/8,3\pi/8)$ points are plotted as functions of $U$.
		The linear system size used here is $L^\prime=8$, which includes these three wave vectors on the Fermi surface, as illustrated in Fig.~\ref{fig:kspace}(f).}
	\label{fig:simultgap}
\end{figure}

\section{\label{app:simult_spgap}SINGLE-PARTICLE GAPS OPEN SIMULTANEOUSLY ON THE ENTIRE FERMI SURFACE}
As depicted in Fig.~\ref{fig:kspace}(f), for the special linear system size $L^\prime=8$, the points $\mathbf{\Sigma}^\prime$ and $\mathbf{\Sigma}^{\prime\prime}$ are situated on distinct Fermi pockets, whereas the $\mathbf{\Sigma}$ point, which is only the contribution from $d_{x^2-y^2}$ orbital, lies at the junction of two Fermi pockets.
The single-particle gaps at these three points, plotted as functions of $U$, are presented in Fig.~\ref{fig:simultgap}.
All three gaps increase monotonically with $U$, and their values are nearly identical.
Given this uniform behavior, it is logical to infer that the entire Fermi surface opens a single-particle gap simultaneously at $U>0$.
Consequently, the single-particle gap at the $\mathbf{\Sigma}$ point is considered a reliable indicator of the system's overall single-particle gap.

\begin{figure}[t]
	\centering
	\includegraphics[width=\linewidth]{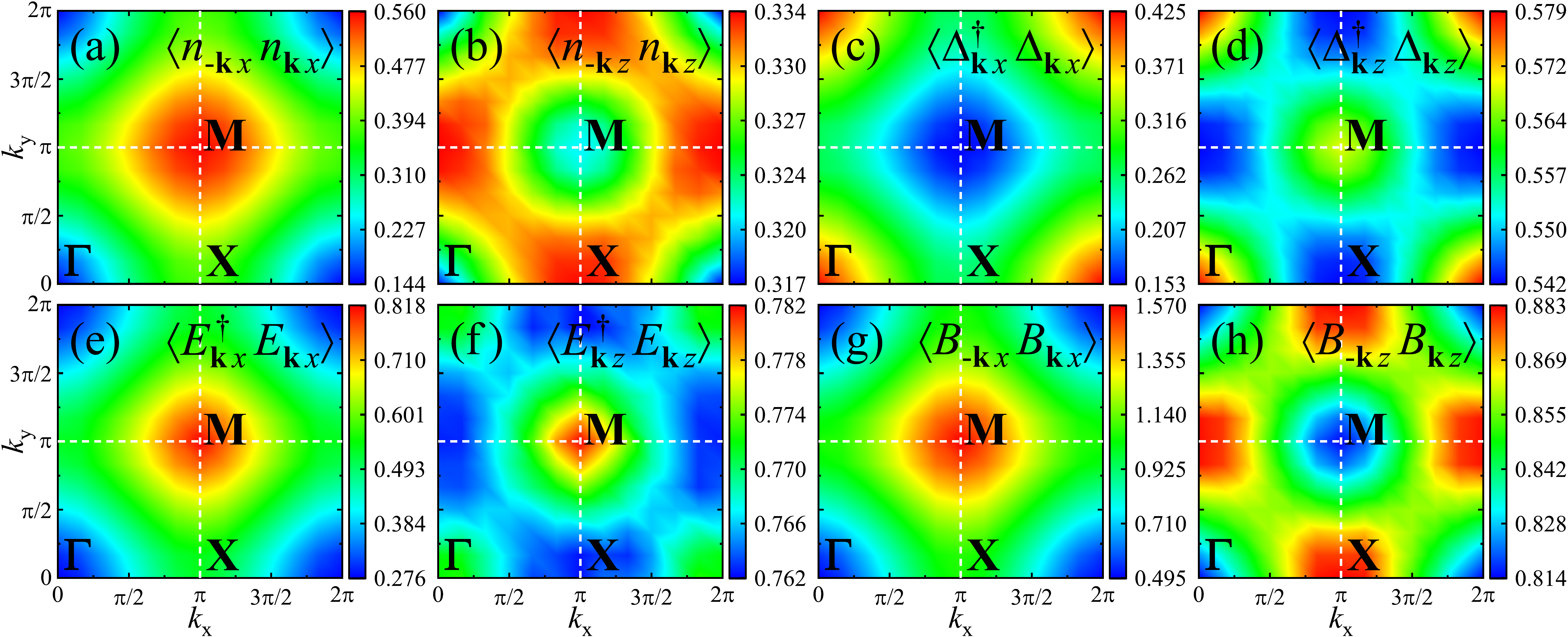}
	\caption{%
		Contour plots of momentum space correlation functions for:
		\mbox{(a)-(b)}~charge-density wave,
		\mbox{(c)-(d)}~pairing,
		\mbox{(e)-(f)}~interlayer exciton, and
		\mbox{(g)-(h)}~bond-order wave at $U/t_1^x=3.0$ and $L=12$.
		The subscripts $x$ and $z$ represent the $d_{x^2-y^2}$ and $d_{3z^2-r^2}$ orbitals, respectively.}
	\label{fig:strfctbz}
\end{figure}

\section{\label{app:absence_conven_order}ABSENCE OF CONVENTIONAL ORDERS IN THE WEAKLY INSULATING PHASE}
Our numerical results indicate that the system is a nonmagnetic insulating phase in the weakly interacting ($0<U<U_c$) region, even with a small finite spin gap.
It is dominated by the inter-layer singlet product state of the $d_{3z^2-r^2}$ orbitals.

\begin{figure}[t]
	\centering
	\includegraphics[width=\linewidth]{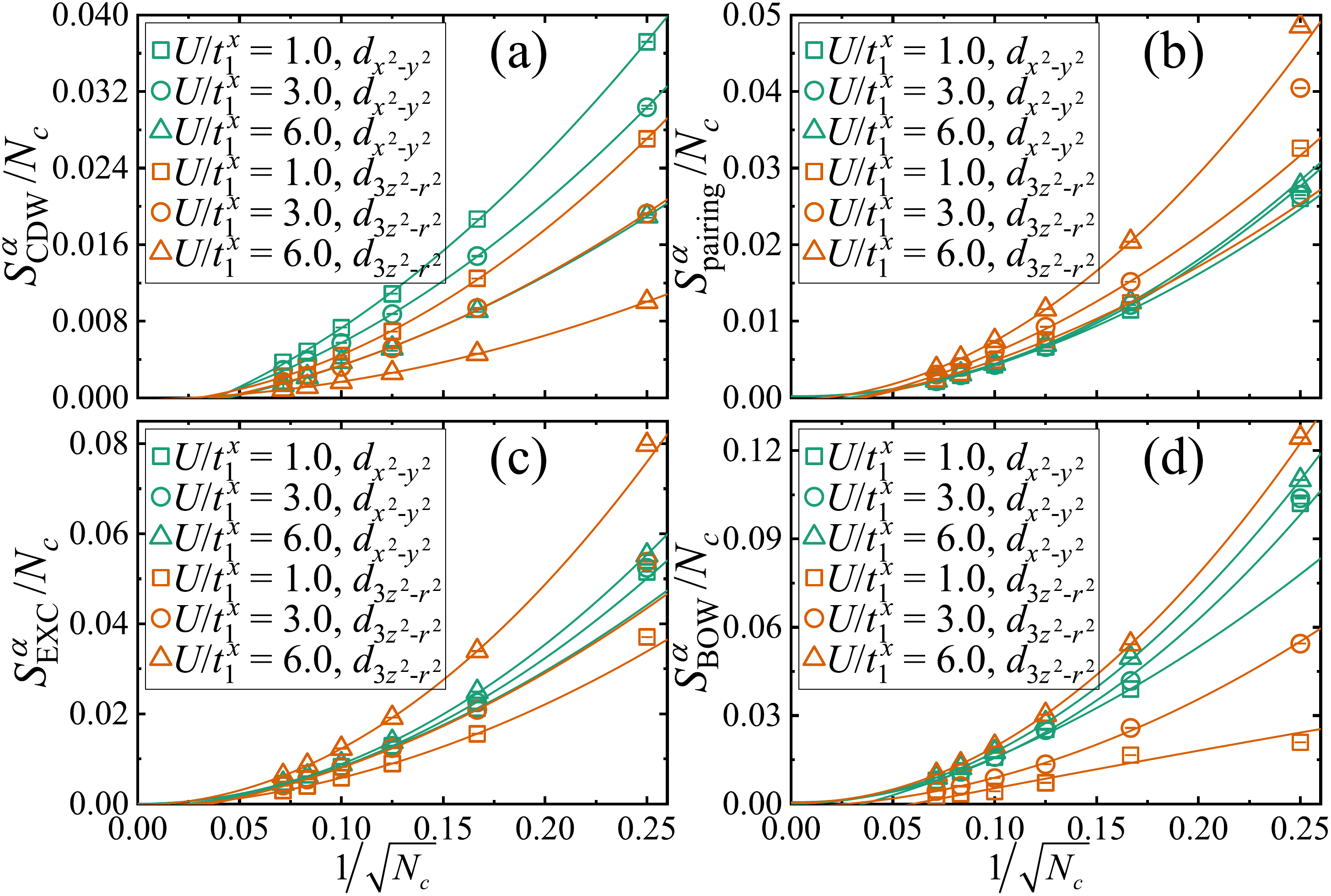}
	\caption{%
		Extrapolations of different structure factors for
		(a)~in-plane charge-density wave,
		(b)~interlayer pairing,
		(c)~interlayer exciton,
		(d)~interlayer bond-order wave.
		They all extrapolate to zero.}
	\label{fig:otheorde}
\end{figure}

To exclude some other conventional orders like charge-density wave (CDW), pairing, interlayer exciton (EXC), and bond-order wave (BOW), we have also calculated their corresponding static-structure factors, which are defined as
\begin{eqnarray*}
	S_\text{CDW}^\alpha(\mathbf{k})&=&\langle n_{-\mathbf{k}\alpha}n_{\mathbf{k}\alpha}\rangle\\
	&=&\frac{1}{2N_c}\sum_{i,j,m}\e^{-\i\mathbf{k}\cdot(\mathbf{R}_j-\mathbf{R}_i)}\\
	&&\times(\langle n_{im\alpha}n_{jm\alpha}\rangle-\langle n_{im\alpha}\rangle\langle n_{jm\alpha}\rangle),\\
	S_\text{pairing}^\alpha(\mathbf{k})&=&\langle\Delta_{\mathbf{k}\alpha}^\dagger\Delta_{\mathbf{k}\alpha}\rangle\\
	&=&\frac{1}{N_c}\sum_{i,j}\e^{-\i\mathbf{k}\cdot(\mathbf{R}_j-\mathbf{R}_i)}\\
	&&\times(\langle\Delta_{i\alpha}^\dagger\Delta_{j\alpha}\rangle-\langle\Delta_{i\alpha}^\dagger\rangle\langle\Delta_{j\alpha}\rangle),\\
	S_\text{EXC}^\alpha(\mathbf{k})&=&\langle E_{\mathbf{k}\alpha}^\dagger E_{\mathbf{k}\alpha}\rangle\\
	&=&\frac{1}{N_c}\sum_{i,j}\e^{-\i\mathbf{k}\cdot(\mathbf{R}_j-\mathbf{R}_i)}\\
	&&\times(\langle E_{i\alpha}^\dagger E_{j\alpha}\rangle-\langle E_{i\alpha}^\dagger\rangle\langle E_{j\alpha}\rangle),\\
	S_\text{BOW}^\alpha(\mathbf{k})&=&\langle B_{-\mathbf{k}\alpha}B_{\mathbf{k}\alpha}\rangle\\
	&=&\frac{1}{N_c}\sum_{i,j}\e^{-\i\mathbf{k}\cdot(\mathbf{R}_j-\mathbf{R}_i)}\\
	&&\times(\langle B_{i\alpha}B_{j\alpha}\rangle-\langle B_{i\alpha}\rangle\langle B_{j\alpha}\rangle),
\end{eqnarray*}
where $n_{im\alpha}=\sum_\sigma n_{im\alpha\sigma}$, $\Delta_{i\alpha}^\dagger=(c_{i1\alpha\uparrow}^\dagger c_{i2\alpha\downarrow}^\dagger-c_{i1\alpha\downarrow}^\dagger c_{i2\alpha\uparrow}^\dagger)/\sqrt{2}$, $E_{i\alpha}^\dagger=c_{i1\alpha\uparrow}^\dagger c_{i2\alpha\uparrow}+c_{i1\alpha\downarrow}^\dagger c_{i2\alpha\downarrow}$, and $B_{i\alpha}=\sum_\sigma(c_{i1\alpha\sigma}^\dagger c_{i2\alpha\sigma}+H.c.)$.
The peak positions of these static-structure factors can be found in Fig.~\ref{fig:strfctbz}; we can use these peaks to do the extrapolations.
However, no CDW, pairing, EXC, and BOW orders are found, as can be seen in Fig.~\ref{fig:otheorde}.
Note that the CDW structure factor here is defined within the plane and the other three order parameters are defined between layers, and they are all defined in the same orbital.
However, for all other possible scenarios, they are also extrapolated to zero.

In addition, we have examined the local spin correlation $\langle\mathbf{S}_{im\alpha}\cdot\mathbf{S}_{jn\lambda}\rangle$ and single-particle correlation $(1/2)\sum_{\sigma}\langle c_{im\alpha\sigma}^\dagger c_{jn\lambda\sigma}+H.c. \rangle$ of finite-size system under open-boundary condition.
But we did not find the intralayer valence bond solid like patterns.

\begin{figure}[b]
	\centering
	\includegraphics[width=0.8\linewidth]{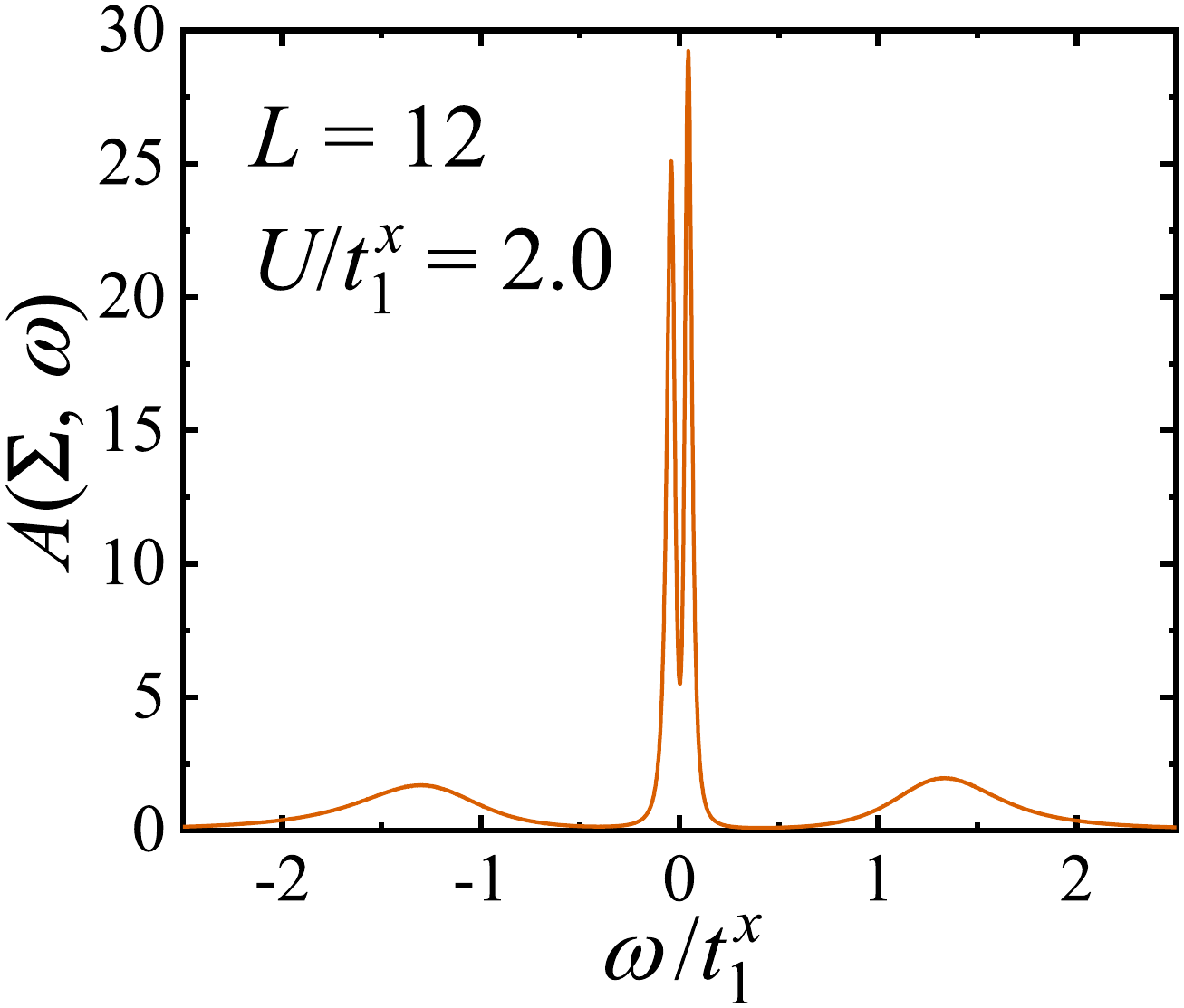}
	\caption{The single-particle spectral function at $\mathbf{\Sigma}$ point for $U/t_1^x=2.0$ and $L=12$.}
	\label{fig:ASigmaOm}
\end{figure}

\section{\label{app:more_sme_reslts}MORE RESULTS OF ANALYTIC CONTINUATION}
The single-particle spectral function at the $\mathbf{\Sigma}$ point for $U/t_1^x=2.0$ and $L=12$ in Fig.~\ref{fig:dynspe}(a) is gapped.
However, due to the small value of this gap and the tendency of the stochastic maximum entropy (SME) algorithm~\cite{assaad2022alf2.0,sandvik1998prb,beach2004arxiv,shao2023physrep} to broaden the spectral function, it is challenging to observe this gap in the contour plot in Fig.~\ref{fig:dynspe}(a) of the main text.
To illustrate this gap, we plot the curve $A(\mathbf{\Sigma},\omega)$ for $U/t_1^x=2.0$ and $L=12$ in Fig.~\ref{fig:ASigmaOm}.
A well-defined local minimum can be observed at the Fermi level $\omega=0$, although its spectral weight does not go to zero completely.
Therefore, there is indeed a small single-particle gap near the Fermi level.
Based on the position of the peak closest to the Fermi level on the left, the single-particle gap can be estimated to be approximately $0.043$, which is very close to $\Delta_\text{sp}/t_1^x\approx0.049$ extracted from $G(\mathbf{\Sigma},\tau)$ in Fig.~\ref{fig:gapextra}(b).

\begin{figure}[b]
	\centering
	\includegraphics[width=\linewidth]{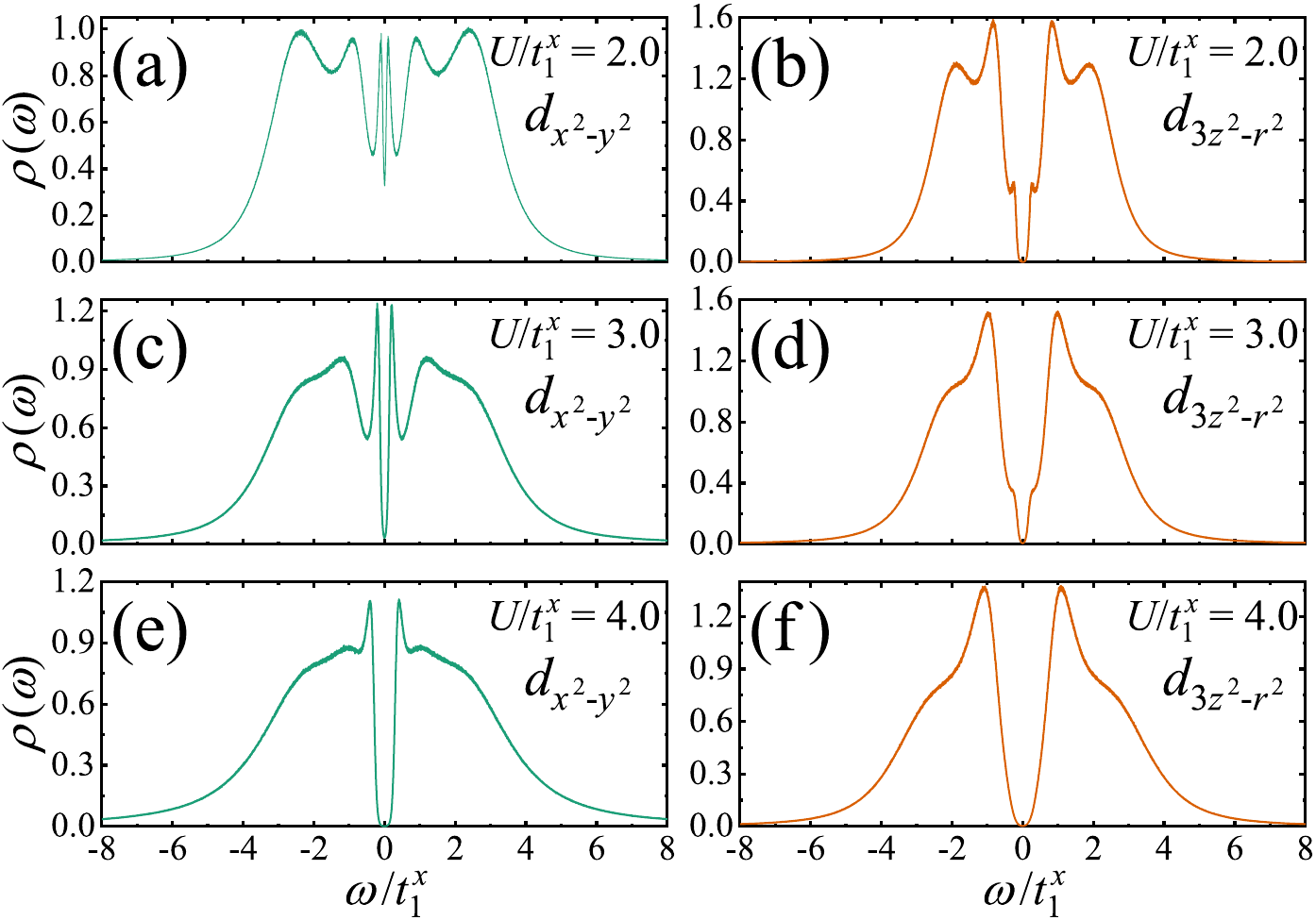}
	\caption{%
		Orbital-resolved local density of states at different values of $U$ with $L=12$.
		(a),
		(c), and
		(e)~are for the $d_{x^2-y^2}$ orbitals, while
		(b),
		(d), and
		(f)~are for the $d_{3z^2-r^2}$ orbitals.}
	\label{fig:RhoOm}
\end{figure}

Moreover, we apply the SME algorithm to
\begin{eqnarray*}
G_\alpha(\mathbf{r}=0,\tau)=\frac{1}{2N_c}\sum_{i,m,\sigma}\langle c_{im\alpha\sigma}^\dagger(\tau)c_{im\alpha\sigma}(0)\rangle,
\end{eqnarray*}
obtaining the orbital-resolved local density of states $\rho(\omega)$ for $L=12$, as shown in Fig.~\ref{fig:RhoOm}.
For the $d_{x^2-y^2}$ orbital, when $U/t_1^x=2.0$, we can still observe a gapped behavior at the Fermi level rather than a typical metallic behavior, as depicted in Fig.~\ref{fig:RhoOm}(a).
As $U$ increases, the gap also gradually increases and becomes well-defined, with the density of states at the Fermi level approaching zero, as displayed in Figs.~\ref{fig:RhoOm}(c) and \ref{fig:RhoOm}(e).
Since the finite-size system with $L=12$ does not encompass any Fermi wave vectors of the $d_{3z^2-r^2}$ orbital [see Fig.~\ref{fig:kspace}(c)], the local density of states of the $d_{3z^2-r^2}$ orbital naturally shows a significant gap which can survive in the thermodynamic limit, as shown in Figs.~\ref{fig:RhoOm}(b), \ref{fig:RhoOm}(d), and \ref{fig:RhoOm}(f).

\bibliography{refs.bib}

\end{document}